\documentstyle[11pt,mrs2001,epsfig]{article}
\begin{document}
\title{The Red-Sequence Cluster Survey: first lensing results}

\author{Henk Hoekstra$^{1,2}$, H. K.C. Yee$^2$, 
and Michael D. Gladders$^2$}
\affil{$^1$ CITA, University of Toronto}
\affil{$^2$ Dept. of Astronomy and Astrophysics, University of Toronto}

\begin{abstract}
The Red-Sequence Cluster Survey (RCS) is a 100 deg$^2$ galaxy cluster
survey designed to provide a large sample of optically selected
clusters of galaxies with redshifts $0.1<z<1.4$. The survey data are
also useful for a variety of lensing studies. Several strong lensing
clusters have been discovered so far, and follow up observations are
underway. In these proceedings we present some of the first results of
a weak lensing analysis based on $\sim$ 24 deg$^2$ of data. We have
detected the lensing signal induced by intervening large scale
structure (cosmic shear) at high significance, and find
$\sigma_8=0.81^{+0.14}_{-0.19}$ (95\% confidence; for a CDM cosmology
with $\Omega_m=0.3,~\Omega_\Lambda=0.7,~h=0.7$). Another application
of these data is the study of the average properties of dark matter
halos surrounding galaxies. We study the lensing signal from
intermediate redshift galaxies with $19.5<R_C<21$ using a
parametrized mass model for the galaxy mass distribution. The
analysis yields a mass weighted velocity dispersion of
$\langle\sigma^2\rangle^{1/2}= 111\pm5$ km/s. In addition we have
constrained for the first time the extent of dark matter halos, and
find a robust upper limit for the truncation parameter $s<470 h^{-1}$
kpc (99.7\% confidence). The biasing properties of these galaxies as a
function of scale are also studied. The RCS data allow us to measure
the ratio of the bias parameter $b$ and the galaxy-mass
cross-correlation coefficient $r$. The results are consistent with a
scale-independent value of $b/r$, for which we find
$b/r=1.05^{+0.12}_{-0.10}$ (for a $\Lambda$CDM cosmology).

\end{abstract}

\section{Introduction}

The Red-Sequence Cluster Survey\footnote{\tt
http://www.astro.utoronto.ca/${\tilde{\ }\!}$gladders/RCS} (e.g.,
Gladders \& Yee 2000) is the largest area, moderately deep imaging
survey ever undertaken on 4m class telescopes. The survey comprises
100 square degrees of imaging in 2 filters (22 widely separated
patches imaged in $R_C$ and $z'$). Ten patches have been observed
using the CFHT 12k mosaic camera, and the remaining 12 southern
patches have been observed using the Mosaic II camera on the CTIO 4m
telescope. The depth of the survey (2 magnitudes past $M^*$ at $z=1$)
is sufficient to find a large number of galaxy clusters to $z\sim
1.4$.

The survey allows a variety of studies, such as constraining cosmological
parameters from the measurement of the evolution of the number density
of galaxy clusters as a function of mass and redshift, and studies
of the evolution of cluster galaxies, blue fraction, etc. at redshifts
for which very limited data are available at present.

The data are also useful for a range of lensing studies. Strong
lensing by clusters of galaxies allows a detailed study of their core
mass distribution. In addition, given the shallowness of the survey,
the arcs are sufficiently bright to be followed up spectroscopically
(e.g., Gladders, Yee, \& Ellingson 2001). Thanks to the large
magnifications of the arcs, it allows unprecedented studies of the
properties of high redshift galaxies. Furthermore, in combination with
detailed modeling of the cluster mass distribution, the geometry of
the images can be used to constrain $\Omega_m$.

Here we concentrate on some of the weak lensing applications, for
which we use 24 deg$^2$ of $R_C-$band survey data (16.4 deg$^2$ of
CFHT, and 7.6 deg$^2$ of CTIO 4m data). We present early results of
the measurement of the lensing signal by intervening large scale
structure (cosmic shear) and a study of the dark matter halos of
(field) galaxies, as well as their biasing properties. The data
analysed so far do not cover complete patches, and therefore we limit
the analysis to the individual pointings.

A detailed description of the data and weak lensing analysis will be
provided in Hoekstra et al. (2001b). The results presented in Hoekstra
et al. (2001b) indicate that the object analysis, and the necessary
corrections for observational distortions work well, which allows us
to obtain accurate measurements of the weak lensing signal.

\begin{figure}[b!]
\begin{center}
\hbox{
\includegraphics[width=8cm,bb=40 195 570 700]{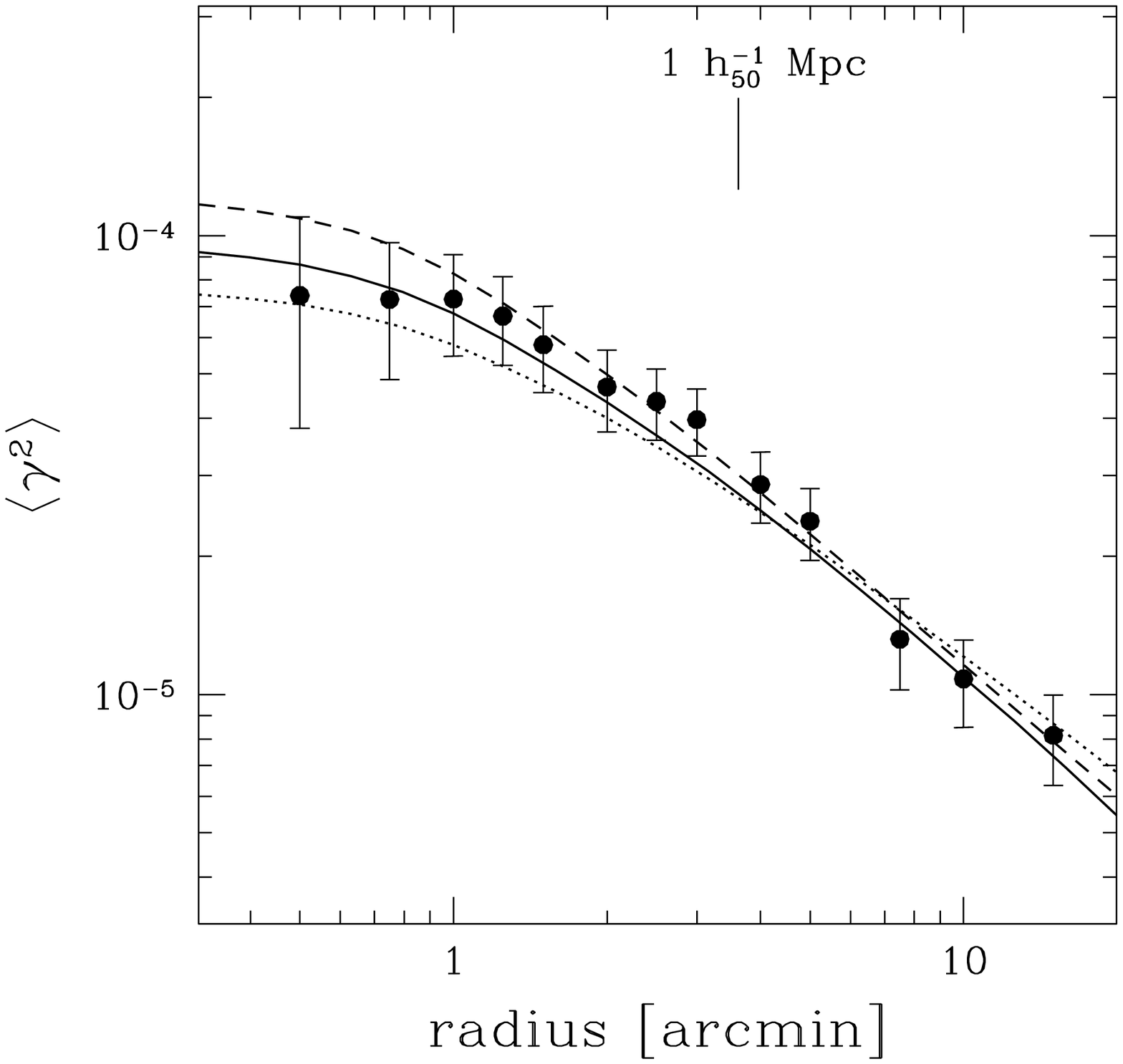}
\includegraphics[height=8cm]{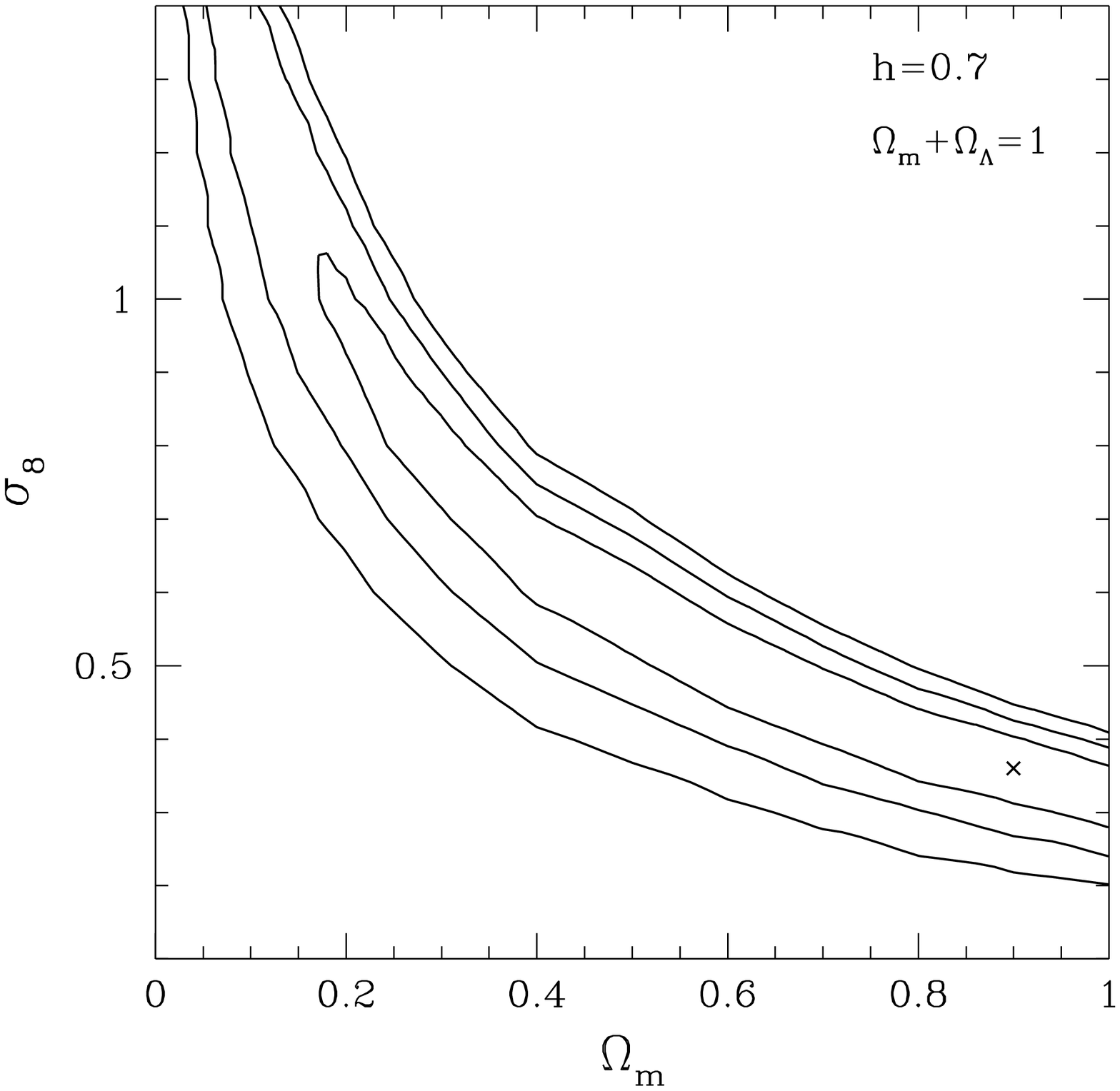}}
\caption{\small {\it left panel}: Measurement of the top-hat smoothed
variance (excess variance caused by lensing by large scale structure)
using galaxies with $20<R_C<24$.  The data consist of 16.4 deg$^2$ of
CFHT data and 7.6 deg$^2$ of CTIO data. The drawn lines correspond to
the expected signals for a SCDM (solid line), OCDM (dashed line), and
$\Lambda$CDM (dotted line) models, using $h=0.7$. The errorbars are
estimated from a large number of realisations of the data set where
the orientations of the galaxies were randomized.  Note that the
points at various scales are strongly correlated.  Under the
assumption that the lensing structures are halfway between the
observer and the sources, a scale of $1~h_{50}^{-1}$ Mpc is indicated.
{\it right panel}: Likelihood contours as a function of
$\Omega_m$ and $\sigma_8$, inferred from the analysis of the top-hat
smoothed variance. The contours have been computed by comparing the
measurements to CDM models with $n=1$, $h=0.7$ and
$\Omega_m+\Omega_\Lambda=1$. The contours indicate the 68.3\%,
95.4\%, and 99.7\% confidence limits on two parameters jointly. Additional
constraints on $\Gamma\approx\Omega_m h$ favour lower values of
$\Omega_m$.
\label{cosmic}}
\end{center}
\end{figure}

\section{Measurement of Cosmic Shear}

The weak distortions of the images of distant galaxies by intervening
matter provide an important tool to study the projected mass
distribution in the universe and constrain cosmological parameters
(e.g., van Waerbeke et al. 2001). Compared to other studies, the RCS
data are relatively shallow, resulting in a lower cosmic shear
signal. However, the redshift distribution of the source galaxies,
which is needed to interpret the results, is known fairly well.  In
addition, our data have been acquired using two different telescopes,
but have been analysed uniformly. A detailed discussion of the
analysis and the results is presented in Hoekstra et al. (2001b).

Comparison of the results from the two telescopes provides a useful
test to check whether the various corrections for observational
distortions have worked well. We find good agreement between the two
measurements, and the combined top-hat smoothed variance is presented
in the left panel of Figure~\ref{cosmic} (16.4 deg$^2$ from CFHT and
7.6 deg$^2$ from CTIO). The signal-to-noise ratio of our measurements
is very good, reaching $\sim 6$ at a radius of 2.5 arcminutes.

We use the photometric redshift distribution inferred from the Hubble
Deep Field North and South to compare the observed lensing signal to
CDM predictions. This redshift distribution works well, as was
demonstrated by Hoekstra, Franx \& Kuijken (2000) for the $z=0.83$
cluster of galaxies MS~1054-03. The right panel in Figure~\ref{cosmic}
shows the inferred likelihood contours for a $\Lambda$CDM cosmology,
with $h=0.7$. For an $\Omega_m=0.3$ flat model we obtain
$\sigma_8=0.81^{+0.14}_{-0.19}$ (95\% confidence), in good
agreement with the measurements of van Waerbeke et al. (2001).

\begin{figure}[b!]
\begin{center}
\hbox{
\includegraphics[height=7.5cm]{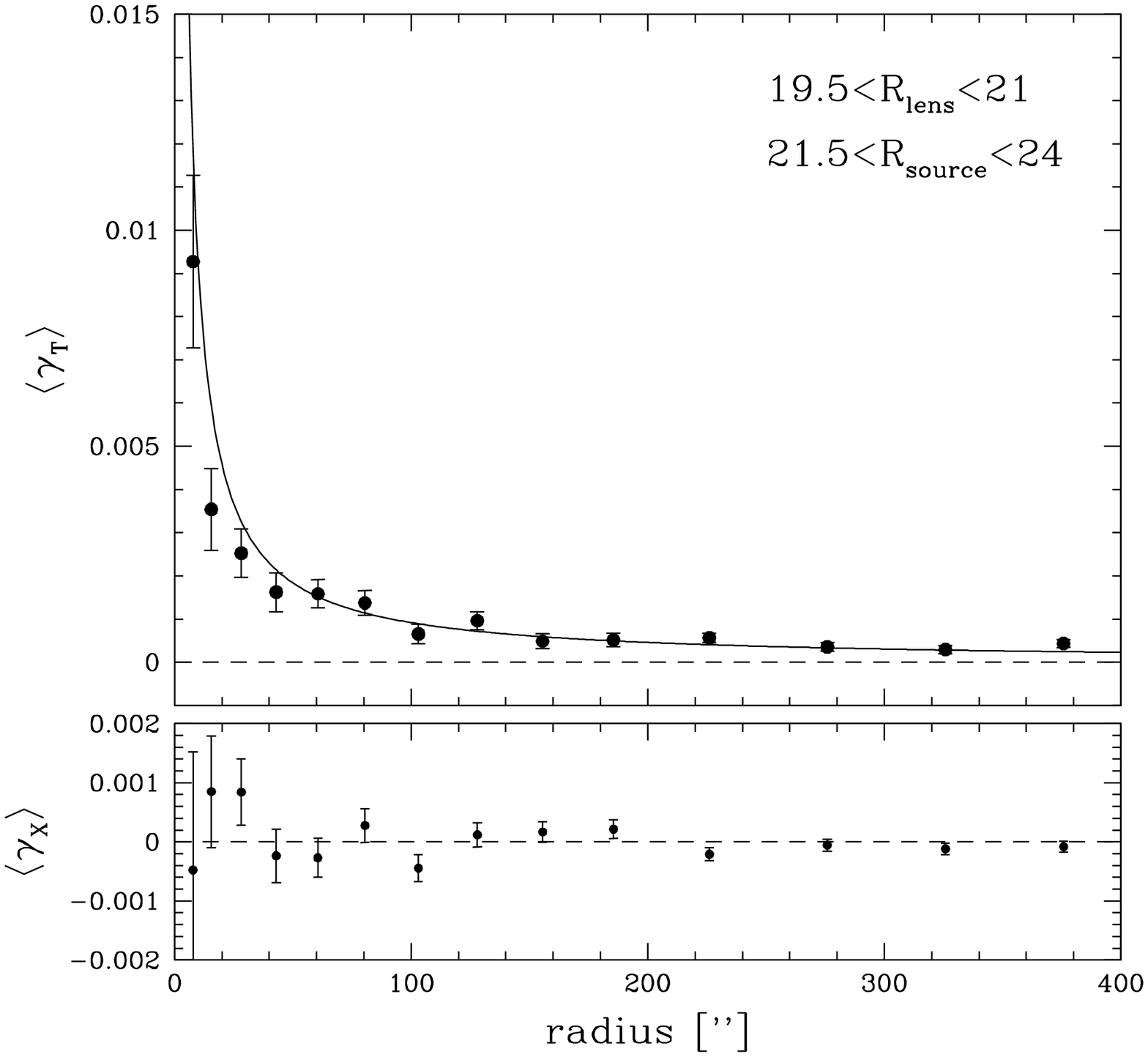}
\includegraphics[height=7.5cm]{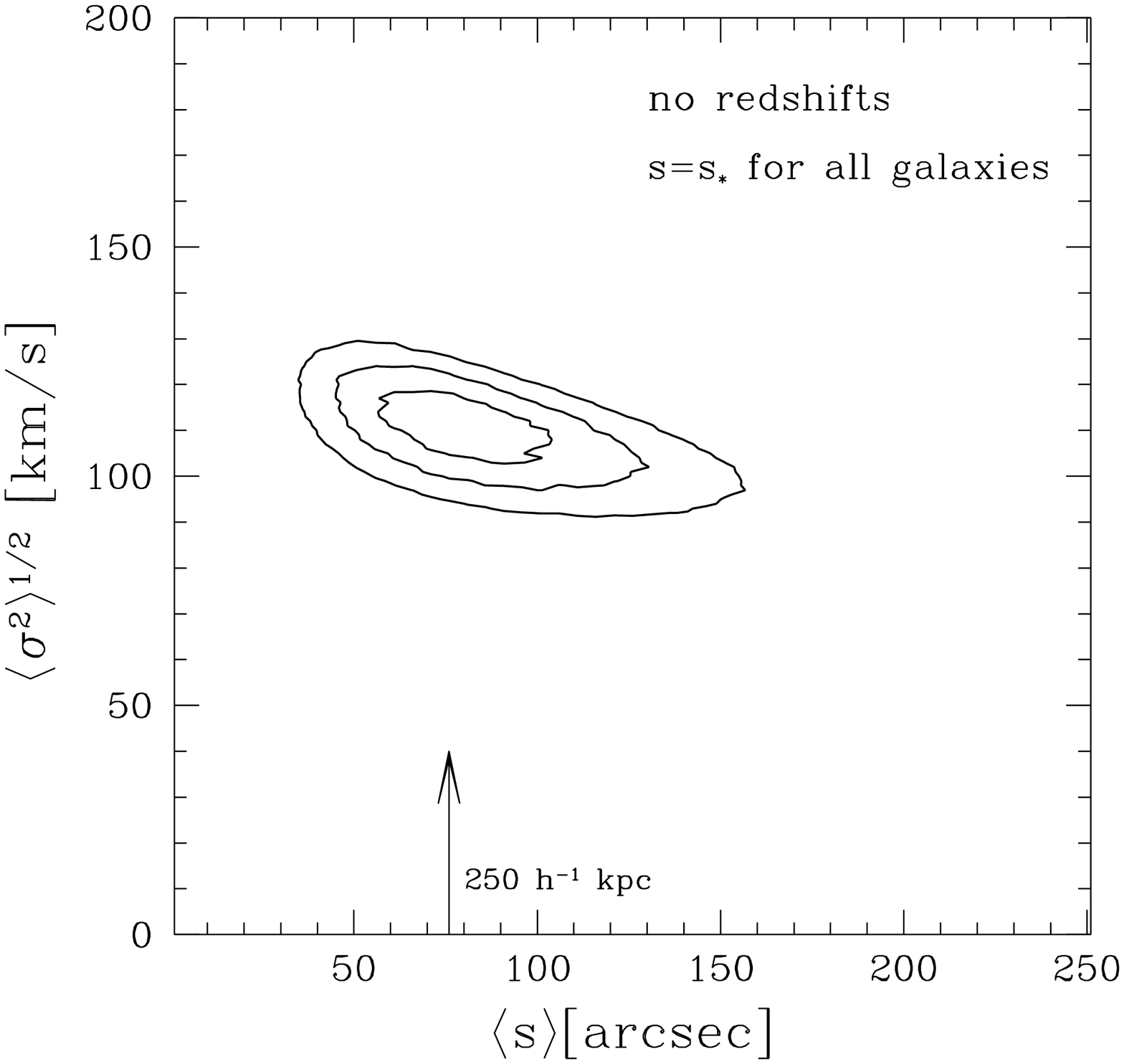}}
\caption{\small {\it left panel}: The ensemble averaged tangential
distortion around galaxies with $19.5<R_C<21$. The solid line
corresponds to the best fit SIS model, for which we find an Einstein
radius $r_E=0.184\pm0.011$ arsec. The lower panel shows the average
signal when the sources are rotated by $\pi/4$. No signal should be
present if the signal in the upper panel is caused by lensing. 
{\it right panel}: Likelihood contours for the mass
weighted velocity dispersion, and the average value of the truncation
parameter $s$. We have also indicated the physical scale when $s$ is
the same for all galaxies. The contours indicate 68.3\%, 95.4\%, and
99.7\% confidence limits on two parameters jointly. For the first
time, we find good constraints on the extent of the dark matter halos
around field galaxies.
\label{galgal}}
\end{center}
\end{figure}

\section{Galaxy-galaxy lensing}

Weak lensing is also an important tool to study the dark matter halos
of field (spiral) galaxies (e.g., Brainerd, Blandford, \& Smail 1996;
Fischer et al. 2000). Rotation curves of spiral galaxies have provided
important evidence for the existence of dark matter halos, but are
confined to the inner regions, as are strong lensing studies. The weak
lensing signal, however, can be measured out to large projected
distances, and it provides a powerful probe of the gravitational
potential at large radii.  Unfortunately, the lensing signal induced
by an individual galaxy is too low to be detected, and one can only
study the ensemble averaged signal around a large number of lenses.

The results presented here are based on 16.4 deg$^2$ of CFHT data.
We use galaxies with $19.5<R_C<21$ as lenses, and galaxies with
$21.5<R_C<24$ as sources which are used to measure the lensing
signal. This selection yields a sample of 36226 lenses and $\sim
6\times 10^5$ sources. The redshift distribution of the lenses is
known spectroscopically from the CNOC2 field galaxy redshift survey
(e.g., Yee et al. 2000), and for the source redshift distribution we
again use the photmetric redshift distribution from the HDF North and
South.  The adopted redshift distributions give a median redshift
$z=0.35$ for the lens galaxies, and $z=0.53$ for the source galaxies.

The ensemble averaged tangential distortion around galaxies with
$19.5<R_C<21$ is presented in the left panel of Figure~\ref{galgal}.
The solid line corresponds to the best fit SIS model, for which we
find an Einstein radius $r_E=0.184\pm0.011$ arcsec. 

A better way to study the lensing signal is to compare the predicted
shear (both components) from a parametrized mass model to the
data. We use the truncated halo model from Schneider \& Rix
(1997). The results are presented in the right panel of
Figure~\ref{galgal}. With the adopted redshift distribution we obtain
$\langle\sigma^2\rangle^{1/2}=111\pm5$ km/s. It turns out that the
quoted value is close to that of an $L_*$ galaxy, and our results are
in fair agreement with other estimates. 

In addition, for the first time, the average extent of the dark matter
halo has been measured. Under the assumption that all halos have the
same truncation parameter, we find a 99.7\% confidence upper limit of
$\langle s\rangle <470 h^{-1}$ kpc.  More realistic scaling relations
for $s$ give lower values for the physical scale of $\langle
s\rangle$, and therefore the result presented here can be interpreted
as a robust upper limit.

We note that the results indicate a steepening of the tangential shear
profile at large radii (which we interpret as a truncation of the halo),
and alternative theories of gravity, such as MOND need to reproduce
this.  A plausible description of lensing by MOND was put forward by
Mortlock \& Turner (2001), which suggests that the shear in MOND drops
$\propto 1/r$.  A preliminary analysis excludes this prediction at high
confidence. 

\section{Measurement of galaxy biasing}

The study of the correlation between the galaxy distribution and the
dark matter distribution (i.e. galaxy biasing) can provide useful
constraints on models of galaxy formation. Most current constraints
come from dynamical measurements that probe relatively large scales
($\ge$ a few Mpc). Schneider (1998) proposed a method that provides
the unique opportunity to study galaxy biasing as a function of scale
using weak lensing. This has been studied in more detail by Van
Waerbeke (1998) who concluded that the results depend only slightly on
the assumed power spectrum of density fluctuations.

In the standard, deterministic linear bias theory, the galaxy density
contrast $\delta_{\rm g}$ is related to the mass density contrast
$\delta$ as $\delta_{\rm g}=b \delta$. However, the biasing relation
need not be deterministic, but might be stochastic. In this case the
galaxy-mass cross-correlation coefficient $r$ is less than 1. Hence,
we allow for stochastic biasing, and include the parameter $r$. 

\begin{figure}[b!]
\begin{center}
\hbox{
\includegraphics[height=7.5cm]{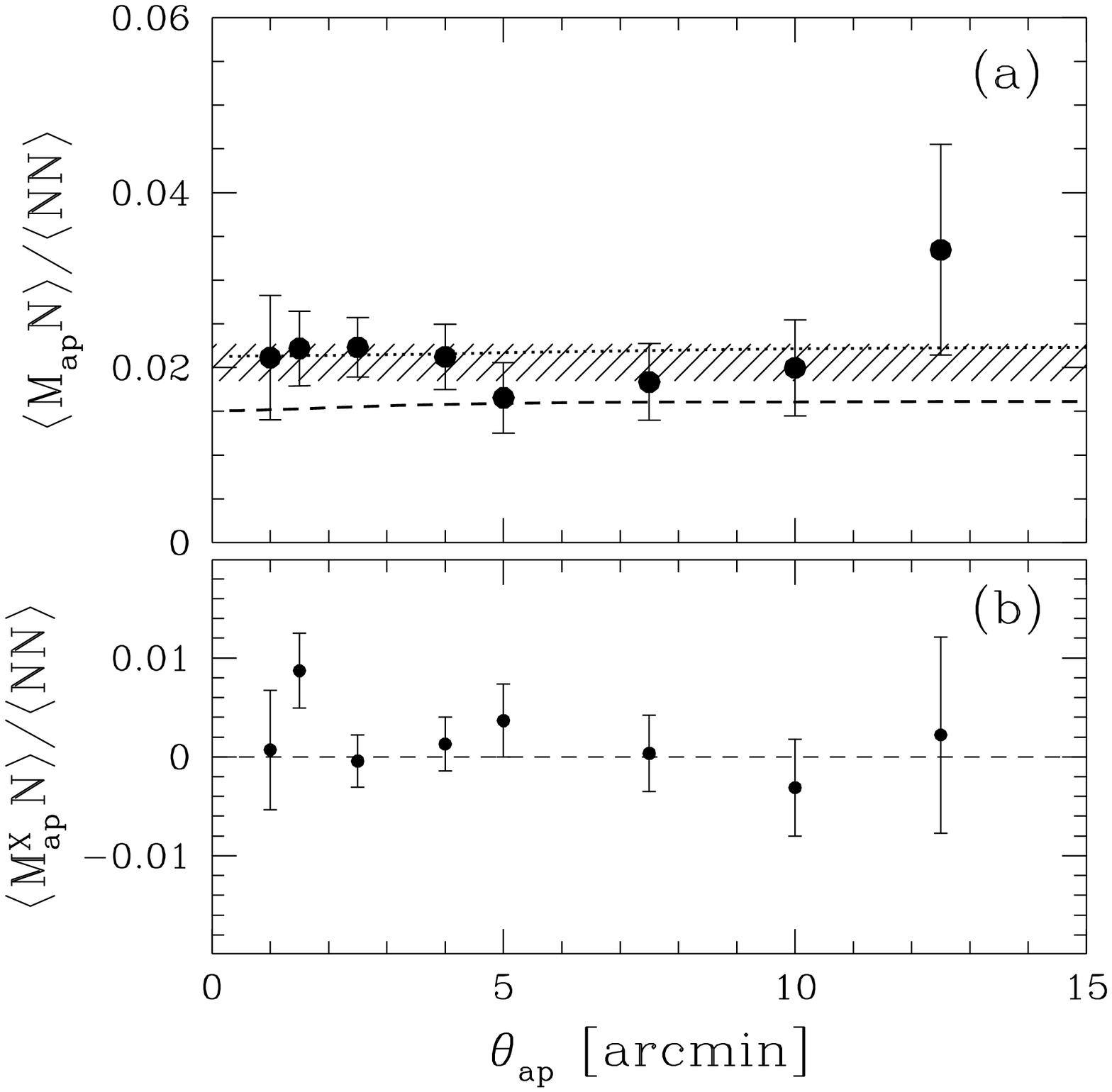}
\includegraphics[height=7.5cm,bb=50 280 590 720]{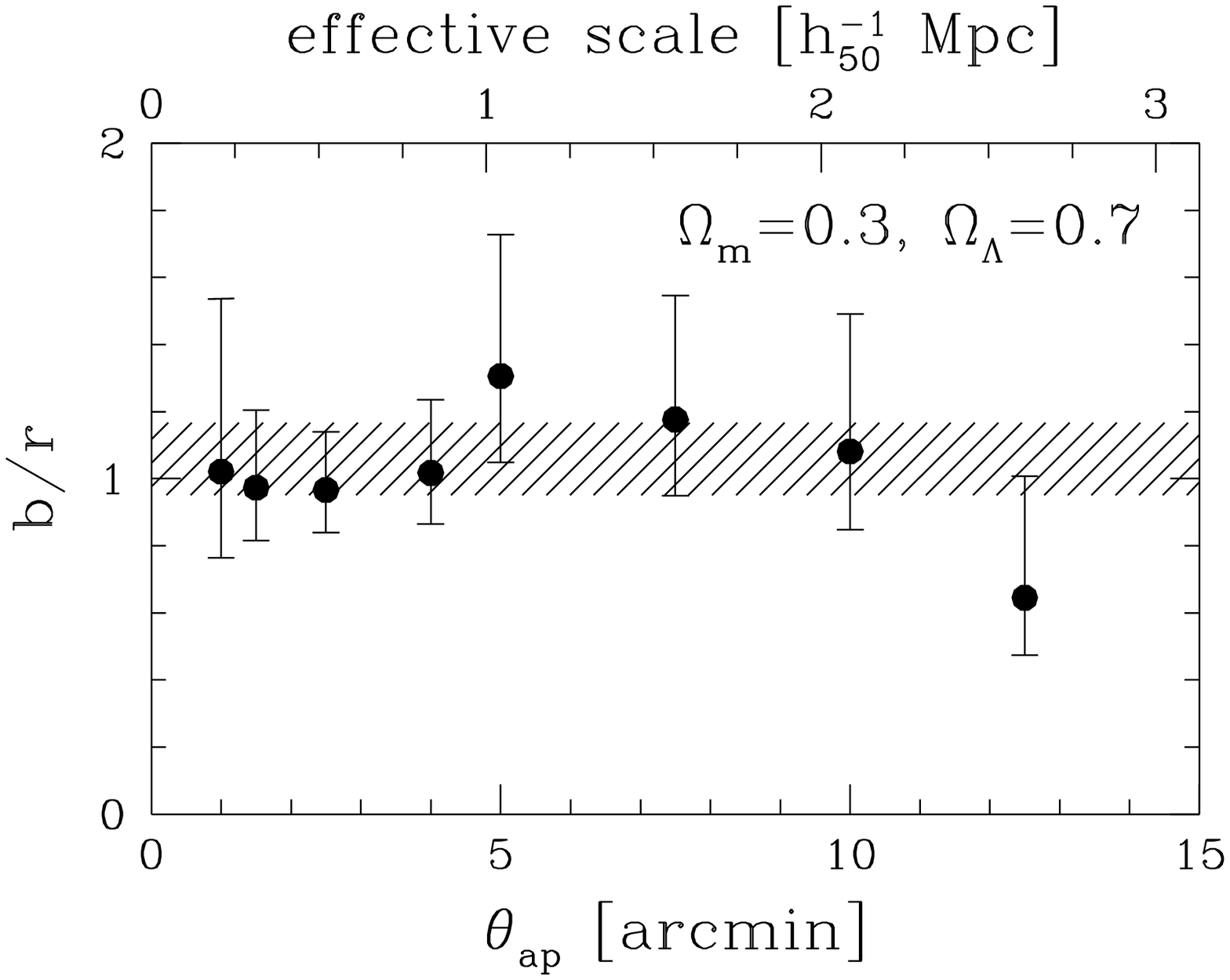}}
\caption{\small {\it left panel}: (a) The observed ratio of the
galaxy-mass cross-correlation function $\langle M_{\rm ap}{\cal
N}\rangle$ and the galaxy auto-correlation function $\langle{\cal
N}^2\rangle$. The dashed line indicates the prediction for an OCDM
model with $b/r=1$, and the dotted line corresponds to a $\Lambda$CDM
model with $b/r=1$. (b) The measured signal when the phase of the
shear is increased by $\pi/2$, which should vanish if the signal in
(a) is caused by lensing. {\it right panel} : Value of $b/r$ as
a function of scale, under the assumption $\Omega_m=0.3$ and 
$\Omega_\Lambda=0.7$. The upper axis indicates the physical
scale corresponding probed. The results are consistent with
a value of $b/r$ that is constant with scale. For this
cosmology we obtain $b/r=1.05^{+0.12}_{-0.10}$ (indicated by the
hatched region). Note that the points are slightly correlated.
\label{bias}}
\end{center}
\end{figure}

We use the 16 deg$^2$ of CFHT imaging data, which allows us to measure
the mass-galaxy cross-correlation and the galaxy auto-correlation function.
Unfortunately the data do not allow a accurate measurement of the
(dark) matter auto-correlation function. A measurement of all 
three correlation functions would enable us to determine both $b$ and
$r$ as a function of scale. In a forthcoming paper we plan to use
the results of van Waerbeke et al. (2001) who measured the 
mass autocorrelation function from deep imaging data. 

Here we present the results based on RCS data only, which results in a
measurement of $b/r$ as a function of scale (for a detailed discussion
see Hoekstra, Yee, \& Gladders 2001a). The ratio of the galaxy-mass
cross correlation and the $\langle M_{\rm ap}{\cal N}\rangle$ and the
galaxy auto-correlation function $\langle{\cal N}^2\rangle$ is
presented in Figure~\ref{bias} (left panel), and the corresponding
estimate of $b/r$ (for a $\Lambda$CDM model) as a function of scale is
shown in the right panel. The results are consistent with a value of
$b/r$ that is independent of scale, in agreement with the prediction
of linear biasing. For the $\Lambda$CDM cosmology we obtain
$b/r=1.05^{+0.12}_{-0.10}$, suggesting that the light distribution
traces the dark matter distribution quite well.

In the near future we will be able to measure the bias parameter
on larger scales, and the accuracy of the estimates will improve
significantly. With additional data this method will allow us
to measure both $b$ and $r$ as a function of galaxy type, scale,
and redshift.

\vfill
\end{document}